\documentclass[a4paper,12pt]{article}

\usepackage{amsmath}
\usepackage[dvips]{graphicx}
\usepackage{cite}

\def\be{\begin{equation}}
\def\eeq{\end{equation}}
\newcommand{\en}{\end{equation}}
\def\ba{\begin{eqnarray}}
\def\ea{\end{eqnarray}}

\def\D{\nabla}
\def\tD{\tilde{\nabla}}
\def\U{\mathcal{U}}
\def\cD{\mathcal{D}}

\makeatletter
\@addtoreset{equation}{section}
\makeatother

\begin{document}

\begin{flushright}
DFTT 21/2005 \\
EPHOU-05-004 \\
July, 2005
\end{flushright}

\vfil

\begin{center}

{\Large Exact Extended Supersymmetry on a Lattice: \\

\vspace{0.3cm}

Twisted $N=2$ Super Yang-Mills in Two Dimensions}\\

\vspace{1cm}

{\sc Alessandro D'Adda}\footnote{dadda@to.infn.it}, 
{\sc Issaku Kanamori}\footnote{kanamori@particle.sci.hokudai.ac.jp}
{\sc Noboru Kawamoto}\footnote{kawamoto@particle.sci.hokudai.ac.jp} and 
{\sc Kazuhiro Nagata}\footnote{nagata@particle.sci.hokudai.ac.jp}\\

\vspace{0.5cm}
{\it{ INFN sezione di Torino, and Dipartimento di Fisica Teorica, Universita 
di Torino, I-10125 Torino, Italy }}\\
and \\
{\it{ Department of Physics, Hokkaido University }}\\
{\it{ Sapporo, 060-0810, Japan}}\\
\end{center}

\vfil

\begin{abstract}
We propose a lattice action for two dimensional super Yang-Mills theory
with a twisted $N=2$ supersymmetry. The extended supersymmetry is fully and
exactly realized on the lattice. The method employed is quite general and its
extension to the $N=4$ supersymmetry in four dimensions is briefly
presented.
The lattice has a new type of ``fermionic'' links, where odd Grassmann
variables, supercharges and fermionic connections sit.
The Leibniz rule is preserved on the lattice, although in a modified
``shifted'' form that takes into account the link nature of both derivatives
and supercharges.
Superfields are semi-local objects and superfield expansion is naturally
embedded in the lattice structure.
The Dirac-K\"ahler twist generates the extended twisted supersymmetry,
 turning the multiplicity of species doublers into the multiplicity due
 to the extended supersymmetry.
In this way the balance
 between bosonic and fermionic degrees of freedom is preserved.
\end{abstract}


\newpage

\setcounter{footnote}{0}

\section{Introduction }

It has been a longstanding problem to formulate exact supersymmetry(SUSY) 
on a lattice. In recent years there was a renewed interest on this problem 
with different approaches\cite{old-lattsusy,D-W-lattsusy,deconstruction,
other-lattsusy1,twist-lattsusy,twist-lattsusy2,other-lattsusy2,
numerical-lattsusy,Leib-fujikawa,Japanese-recent,DKKN}.  We also proposed in
\cite{DKKN} a discrete $N=D=2$ twisted SUSY algebra on a lattice. A chiral and 
anti-chiral pair of superfields reproduced lattice super BF and Wess-Zumino 
actions which have exact twisted SUSY for all supercharges. 
In this article we extend our previous formulation to include gauge fields, and
we write a lattice action for $N=2$ twisted super Yang-Mills(SYM) 
in two dimensions.       
The main new feature of our approach compared to others is that our lattice 
actions have exact discrete twisted SUSY invariance for all supercharges. 
This was not achieved before.
The formulation of the twisted superalgebra by a Dirac-K\"ahler twisting 
procedure was proposed in the continuum for a number of theories, including 
$N=4$ twisted SUSY in four dimensions\cite{KT,KKM}, and it plays a fundamental
role in preserving all supersymmetries on the lattice. 
In fact it is only after the Dirac-K\"ahler twist that it is possible to 
construct a lattice, made of both bosonic and fermionic links, that embeds in
its links the structure of the extended SUSY algebra.
The existence of such a lattice, for both the $N=2$ twisted SUSY in two
dimensions and the $N=4$ twisted SUSY in four dimensions, is a non-trivial result and it
is at the root of the developments described in the present paper.

\section{Discretization of Twisted SUSY Algebra}
One of the difficulties in formulating exact supersymmetry on a
lattice is the breaking of the Poincar\`e group by the lattice.
As a consequence derivatives are replaced on the lattice by finite differences,
and these do not satisfy the Leibniz rule.  This is an important point, and
defining lattice derivatives that satisfy the Leibniz rule is
a crucial step in establishing exact supersymmetry on a lattice.

We will follow essentially the same approach
already presented in ref \cite{DKKN}, but with one important difference:
we will show that the ``mild non-commutativity'' introduced in \cite{DKKN,KK}
just reflects the link nature of the momentum and of the
supersymmetry  operators on the lattice.

The left finite difference operator $\Delta_{+\mu}$ defined by:
\be
\left(\Delta_{+\mu} \Phi\right)(x) = \Phi(x+ n_{\mu}) - \Phi(x)
\label{finite}
\eeq
does not satisfy the usual Leibniz rule, but rather a modified
``shifted'' one:
\be
\left(\Delta_{+\mu} \Phi_1 \Phi_2\right)(x) = \left( \Delta_{+\mu} \Phi_1
 \right)(x) \Phi_2(x) + \Phi_1(x+ n_{\mu}) \left( \Delta_{+\mu}
 \Phi_2\right)(x)
 \label{lrule}
 \eeq
which can be associated to a ``shifted'' commutator:
\be
\left(\Delta_{+\mu} \Phi \right)(x) = \Delta_{+\mu} \Phi(x) -
\Phi(x+n_{\mu})  \Delta_{+\mu} .
\label{shcomm}
\eeq

Eq. (\ref{shcomm}) is best interpreted by considering the operator
$\Delta_{+\mu}$ as a link variable defined on the oriented link
$(x+n_{\mu}, x)$. The shift in the argument of $\Phi$ then arises naturally.
Comparison of (\ref{shcomm}) with (\ref{finite}) shows that
$\Delta_{+\mu}$ is a constant link variable that takes the value $-1$ on
the link $(x+n_{\mu}, x)$ for any value of $x$:
\be
(\Delta_{+\mu})_{x+n_{\mu}, x} = -1 .
\label{minusone}
\eeq
Similarly the right finite difference operator $\Delta_{-\mu}$ can be
defined as a link variable that takes the value $+1$ on the link
$(x-n_{\mu},x)$ for any $x$.

The anti-commutator of two supercharges should on the lattice
give either a left or a right finite difference operator, so it is
natural to assume that
the SUSY transformations generated by a supercharge $Q_A$
 are defined on the lattice by a shifted (anti-)commutator of the type
 (\ref{shcomm}):
\be
 s_A \Phi (x, \theta) = Q_A \Phi (x, \theta) - \Phi ( x + a_A,\theta)
 Q_A
\label{susytrans}
\eeq
where $a_A$ are some new shifts to be defined.
This implies implementing the lattice with fermionic links
$(x + a_A, x)$, the supercharges $Q_A$ sitting on them as
link variables.
Consistently the Grassmann variables $\theta_A$ that define the
superspace describe constant link variables associated to the links
$(x, x+a_A)$, $\theta_A$  being their common value.
Similarly $\frac{\partial}{\partial \theta_A}$ will be associated
to constant link variables on the same links as
$Q_A$, namely $(x + a_A, x)$, and can be viewed as the same link
variable as $\theta_A$ with opposite link orientation.
The standard algebraic properties of the $\theta_A$-$\frac{\partial}
{\partial \theta_A}$ variables should now be interpreted as link relations,
for instance $ \theta_A \theta_B + \theta_B \theta_A= 0$ would read
\be
( \theta_A)_{x,x+a_A} (\theta_B)_{x+a_A,x+a_A+a_B} + (\theta_B)_{x,x+a_B}
(\theta_A)_{x+a_B,x+a_A+a_B}= 0.
\label{theta}
\eeq

The superfield structure is then mapped into the fermionic lattice
structure, with the superfield expansion now reading as follows:
\begin{eqnarray}
 F(x,\theta)
  &=& f(x)+(\theta_A)_{x,x+a_A} f_A(x+a_A) \nonumber \\
  &&  {}+(\theta_A)_{x,x+a_A}(\theta_B)_{x+a_A,x+a_A+a_B}f_{AB}(x+a_A+a_B)
     + \cdots.
\label{supexp}
\end{eqnarray}
The link nature of $\theta_A$ has been emphasized here, but the
link labels in $(\theta_A)_{x,x+a_A}$ may be dropped  as $\theta_A$
is a constant link variable.
The expansion (\ref{supexp}) coincides with the one given in
\cite{DKKN}, however no reference to non-commutativity is needed any
longer as all shifts in the arguments naturally follow from the link
nature of the variables involved.

The non-trivial part of this lattice formulation is finding a set of
shifts $a_A$ that is consistent with the SUSY algebra.
The choice of $a_A$ implies  choosing a base in the
space of the supercharges, namely choosing
the orientation of the fermionic links, in the same way as the choice
of $n_{\mu}$ defines a base in the space of translations and
defines the orientation of the bosonic lattice.
This choice is non-trivial and not without consequences: consistency
with the SUSY algebra is obtained only with very specific
choices of the base and of the shifts $a_A$, and this choice may or may not 
break  R-symmetry and the residual discrete Lorentz invariance.
Preserving R-symmetry and Lorentz invariance is a delicate and
important problem that will not be addressed in the present paper.

Let $Q_A$ be the supercharges of the chosen base and $a_A$ the corresponding
shifts. Consistency  with the SUSY algebra
requires  that the only non-vanishing anti-commutators of two
supercharges give the  finite difference operator in
one of the lattice directions, namely for some $A$, $B$ and $\mu$:
\be
\{Q_A,Q_B\} = \Delta_{\pm \mu}
\label{cons}
\eeq
and 
\be
a_A + a_B = \pm n_{\mu}.
\label{shifts}
\eeq
To our knowledge there are only two SUSY algebras where
conditions (\ref{cons}) and (\ref{shifts}) can be satisfied 
for some $Q_A$ and $a_A$: the $N=2$ twisted supersymmetry
in two dimensions and the $N=4$ twisted supersymmetry in four dimensions.

The solution for the two dimensional $N=2$ twisted SUSY algebra was found
already in \cite{DKKN}. The key point, as mentioned before, is the choice
of the correct base in the space of the SUSY generators
$Q_{\alpha i}$ ( $\alpha$ and $i$ both take the values $1$ and $2$
and are respectively the Lorentz spinor index and the internal
symmetry index labeling the two different $N=2$ supercharges).
The only base that makes the lattice discretization of the
SUSY algebra consistent in the sense of Eq.(\ref{cons}) and
(\ref{shifts}) is obtained by treating the two indices in $Q_{\alpha i}$
as matrix indices and expanding $Q_{\alpha i}$ on the base of
two dimensional $\gamma$ matrices:
\be
Q_{\alpha i} = \left( \mathbf{1} Q + \gamma^{\mu} Q_{\mu} +
\gamma^5 \tilde{Q} \right)_{\alpha i}.
\label{gammaex}
\eeq
The non-vanishing anti-commutators of the discretized twisted SUSY algebra
then read:
\be
\{ Q , Q_{\mu} \} = i \Delta_{+\mu}, \qquad
\{\tilde{Q}, Q_{\mu} \} = -i \epsilon_{\mu \nu} \Delta_{- \nu}
\label{discrsusy}
\eeq
with the shifts $a_A$ satisfying the following relations:
\be
a + a_{\mu} = n_{\mu},~~~~~~~\tilde{a}+ a_{\mu}=
-|\epsilon_{\mu \nu}| n_{\nu},~~~~~~~a+\tilde{a}+a_1+a_2=0.
\label{as}
\eeq

\begin{figure}[b]
\hfil \includegraphics[width=35mm]{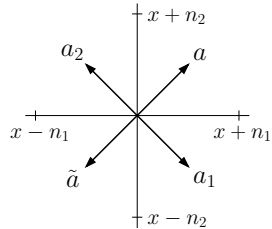}
\caption{Symmetric choice of shift parameter $a_{A}$}
\label{symmetric_a}
\end{figure}

One shift (for instance $a$) is not determined by (\ref{as}) and can
be chosen arbitrarily. 
A convenient choice, that will be often used in what follows and is shown in
Fig.~\ref{symmetric_a}, is the symmetric choice where $a=(1/2,1/2)=-\tilde{a}$ 
and $a_1=-a_2=(1/2,-1/2)$.

It is important to realize that $Q_{\alpha i}$ is identified as $N=2$ extended 
supercharges, satisfying 
\be
  \{Q_{\alpha i},\overline{Q}_{j \beta}\} = 
2i \delta_{ij} (\gamma)_{\alpha\beta}\partial_\mu, 
\label{sp-cha-atcom}
\eeq 
where $ \overline{Q}_{i\alpha} =(C^{-1}Q^TC)_{i\alpha} $. In Euclidean two 
dimensions we can take $C={\bf 1}$ for Majorana representation. 

Although not explicitly shown in (\ref{discrsusy}) all anti-commutators in
it must be understood as link relations according to the previous
discussion, the conditions (\ref{as}) ensuring the consistency.
For instance the explicit form of the first equation in (\ref{discrsusy})
showing the link structure is:
\be
\{Q,Q_{\mu}\}_{x+n_{\mu},x}=(Q)_{x+n_{\mu},x+a_{\mu}}(Q_{\mu})_{x+a_{\mu},x}+
(Q_{\mu})_{x+n_{\mu},x+a}(Q)_{x+a,x} = i (\Delta_{+\mu})_{x+n_{\mu},x}.
\label{explicit}
\eeq
The  form of the supercharges in terms of the
$\theta_A$ variables
was given in \cite{DKKN}. With the link indices explicitly showing they read:
\begin{align}
(Q)_{x,x-a}&=\left( \frac{\partial}{\partial \theta}\right)_{x,x-a} +
\frac{i}{2} (\theta^\mu)_{x,x+a_\mu} (\Delta_{+\mu})_{x+a_\mu,x-a}, \nonumber \\
 (\tilde{Q})_{x,x-\tilde{a}}&=\left( \frac{\partial}{\partial
 \tilde{\theta}}\right)_{x,x-\tilde{a}} -
\frac{i}{2}\epsilon_{\mu \nu} (\theta^\mu)_{x,x+a_\mu}
(\Delta_{-\nu})_{x+a_\mu,x-\tilde{a}}, \label{susych}
\\
(Q_{\mu})_{x,x-a_\mu}&=\left( \frac{\partial}{\partial \theta_\mu }
\right)_{x,x-a_\mu}+ \frac{i}{2} (\theta)_{x,x+a} (\Delta_{+\mu})_{x+a,x-a_\mu}
 - \frac{i}{2} \epsilon_{\mu \nu} (\tilde{\theta})_{x,x+\tilde{a}} (\Delta_{-\nu})
 _{x+\tilde{a},x-a_\mu}.\nonumber
\end{align}
The covariant derivatives $D$, $\tilde{D}$ and $D_{\mu}$ are defined in the 
same way but with a change of sign  with respect to (\ref{susych}) in front of
the terms containing $\Delta_{\pm \mu}$.

A discretized $N=4$ SUSY algebra in four dimensions can be
found following exactly the same procedure as for the $N=2$ in two
dimensions: first we expand the $16$ supercharges $Q_{\alpha
i }$ in the ``twisted'' base of the four dimensional $\gamma$ matrices:
\be
Q_{\alpha i}= \frac{1}{\sqrt{2}} \left(\mathbf{1} Q + \gamma^\mu
Q_\mu + \frac{1}{2} \gamma^{\mu \nu} Q_{\mu \nu} + \tilde{\gamma}^\mu
\tilde{Q}_{\mu} + \gamma_5 \tilde{Q} \right)_{\alpha i}.
\label{q4}
\eeq
It is again important to realize that a conjugate charge can be defined 
just as in the two dimensional case and satisfies an extended $N=4$ super 
algebra where $C=-C^T$ in Euclidean four dimensions. The details of the 
notation and the full $N=4$ twisted SUSY algebra by Dirac-K\"ahler twist 
will be found in \cite{KKM}.

The discretized twisted SUSY algebra can then be written as
(only non-vanishing anti-commutators are given):
\ba
\{Q,Q_\mu \} &=& +i\Delta_{+\mu},~~~~\{Q_\mu,Q_{\rho \sigma}\}=
-i\delta_{\mu \rho} \Delta_{-\sigma} +i \delta_{\mu \sigma}
\Delta_{-\rho}, \nonumber \\
\{\tilde{Q},\tilde{Q}_\mu \} &=& +i\Delta_{-\mu},~~~~\{\tilde{Q}_\mu,
Q_{\rho \sigma}\}=+i\epsilon_{\mu\rho\sigma\nu}\Delta_{+\nu}.
\label{4susy}
\ea
All charges $Q_A$ are link variables defined on links $(x,x-a_A)$
and the anti-commutators are to intend a link anti-commutators.
The consistency conditions for the shifts $a_A$ are given by:
\ba
a+a_\mu=n_\mu,~~a_\mu+a_{\mu\rho}=-n_\rho,~~
\tilde{a}_{\mu}+a_{\rho\sigma}=|\epsilon_{\mu\nu\rho\sigma}|n_{\nu},
~~\tilde{a}+\tilde{a}_{\mu}=-n_\mu,
\label{4a}
\ea
for $\mu\neq\rho\neq\sigma$.
Again one shift is not determined by (\ref{4a}).

We are not going to develop the four dimensional $N=4$
theory in this paper. However the very existence of the algebra (\ref{4susy})
with the consistency conditions (\ref{4a}) is important as it denotes
that a lattice formulation of the $N=4$ SYM with all
supersymmetry exact is a likely possibility.

Both in the $N=2$ and in the $N=4$ case the lattice formulation requires
that the supercharges are expanded in the ``twisted'' base.
This is certainly not accidental, and its deep reason resides in  the chiral 
fermion problem. In fact the same type of  expansion for a fermionic field
$\psi_{\alpha i}(x)$ gives rise to Dirac-K\"ahler fermions, which then
are the only one to have a precise correspondence with the geometrical
elements of the lattice (sites, links, etc.). 
Dirac-K\"ahler fermions may also be interpreted as fermion doublers, thus
trading the multiplicity due to extended supersymmetry for the
multiplicity due to the doubling phenomenon. 
The notorious ``flavor'' degrees of freedom originated from species doublers 
is now identified as extended SUSY degrees of freedom, which is stressed 
in \cite{DKKN,KT,KKM}.
In this way the balance
between bosonic and fermionic degrees of freedom is not affected by
doubling and supersymmetry can be preserved. This is ultimately the
reason why only extended supersymmetries can be discretized, at least in
the present approach.

In ref.~\cite{DKKN} we used the discrete $N=2$ algebra (\ref{discrsusy}) to
formulate supersymmetric $N=2$ BF theory and Wess-Zumino model on the
lattice keeping all supersymmetries exact, the ``mild
non-commutativity'' introduced there being equivalent to the present link
variable formulation. We want now to discuss the case of supersymmetric
gauge theories, beginning with the $N=2$ SYM theory in two
dimensions and leaving the more difficult case of the four dimensional
$N=4$ SYM theory to future investigation.
In this context the present link interpretation proves very useful, in
fact in order to introduce the gauge variables it is enough to replace
the constant link variables introduced above (both bosonic and
fermionic) with corresponding gauge degrees of freedom:

\be
\Delta_{\pm \mu}  \rightarrow  \mp \mathcal{U}_{\pm \mu},~~~~~~~
Q_A \rightarrow \nabla_A 
\label{corresp}
\eeq
where $ \mathcal{U}_{\pm \mu}$ and $\nabla_A$ are $x$ dependent
\be
 ( \mathcal{U}_{\pm \mu})_{x \pm n_\mu,x} = \mathcal{U}_{x \pm
 n_\mu,x}, ~~~~~~~~~~(\nabla_A)_{x+a_A,x} = \nabla_{x+a_A,x}
\label{corresp2}
\eeq
and carry indices of the gauge group (say $SU(N)$). 
In principle in this correspondence  also the fermionic superspace
coordinates $\theta_A$ and the corresponding derivatives
$\frac{\partial}{\partial \theta_A}$ should  become gauge link
variables, and one should be able to express $\nabla_A$ in terms of them
and of  $ \mathcal{U}_{\pm \mu}$. However all we need to formulate the
theory is the algebra of $\nabla_A$ and $\mathcal{U}_{\pm \mu}$, so this
problem does not need to be addressed yet\footnote{Nor we need to address here
the role of the covariant derivatives $D_A$ in this correspondence as the
definition of chiral superfields is not required.}.
In the lattice formulation of $N=2$ SYM theory the link variables 
$ \mathcal{U}_{x \pm  n_\mu,x} $ are not unitary matrices:
 $ \mathcal{U}_{x+n_\mu,x}\,\mathcal{U}_{x,x+n_\mu}\neq 1 $. 
 In fact if one looks at the theory as resulting from dimensional reduction of
 the $N=1$ theory in four dimension one finds that the gauge link variables
 $\U_{\pm\mu}$ may be expressed as
\begin{eqnarray}
(\U_{\pm\mu})_{x\pm n_{\mu},x} 
= (e^{\pm i(A_{\mu}\pm i\phi^{(\mu)})})_{x\pm n_{\mu},x},
\label{def-of-glink}
\end{eqnarray}
where the non-unitary parts $\phi^{(\mu)}(\mu=1,2)$ are the remnants of the 
gauge fields in the two compactified dimensions and represent the two
scalar fields of the compactified theory.

\section{Lattice Formulation of Twisted $N=2$ Super Yang-Mills in Two Dimensions}

Based on the arguments given in the previous sections we explicitly 
construct $N=D=2$ super Yang-Mills (SYM) action which has an exact lattice 
SUSY invariance for all twisted supercharges 
in a manifestly gauge covariant manner. The formulation we propose here
essentially corresponds to a lattice counterpart of the superconnection 
formalism given in \cite{KKM}. In contrast with the continuum formulation, 
we need not explicitly refer to superconnections and Wess-Zumino gauge fixing 
in the lattice formulation. 
The main ingredients are gauge covariant bosonic and fermionic link 
variables $\U_{\pm\mu}$ and $\D_{A}$ which carry the same shifts as 
$\Delta_{\pm\mu}$ and $Q_A$, as summarized in Table~\ref{shifting_nature}.
The gauge transformation of link variables on a lattice are  given by 
\begin{eqnarray}
(\U_{\pm\mu})_{x\pm n_{\mu},x} &\rightarrow&
G_{x\pm n_{\mu}}(\U_{\pm\mu})_{x\pm n_{\mu},x}G^{-1}_{x},\\
(\D_{A})_{x+a_{A},x} &\rightarrow& G_{x+a_{A}}(\D_{A})_{x+a_{A},x}G^{-1}_{x},
\end{eqnarray}
where $G_{x}$ denotes the finite gauge transformation at the site $x$.
Next we impose the following $N=2$ SYM constraints 
on a lattice,
\begin{eqnarray}
\{\D,\D_{\mu}\}_{x+a+a_{\mu},x} &=& + i (\U_{+\mu})_{x+n_{\mu},x}, 
\label{gauge1} \\[2pt]
\{\tD,\D_{\mu}\}_{x+\tilde{a}+a_{\mu},x} 
&=& +i\epsilon_{\mu\nu}\ (\U_{-\nu})_{x-n_{\nu},x}, \label{gauge2}\\[2pt]
\{others\} &=& 0,
\label{gauge3}
\end{eqnarray}
where  $\U_{\pm\mu}$ are the gauge link variables introduced in (\ref{corresp}) and
defined in (\ref{def-of-glink}). 
 The left hand side (l.h.s.) of (\ref{gauge1})--(\ref{gauge3}) should be 
understood as link  anti-commutators
\begin{align}
\{\D,\D_{\mu}\}_{x+a+a_{\mu},x} &\ =\  (\D)_{x+a+a_{\mu},x+a_{\mu}}
(\D_{\mu})_{x+a_{\mu},x}
+(\D_{\mu})_{x+a+a_{\mu},x+a}(\D)_{x+a,x},\\[2pt]
\{\tD,\D_{\mu}\}_{x+\tilde{a}+a_{\mu},x} 
&\ =\ (\tD)_{x+\tilde{a}+a_{\mu},x+a_{\mu}}
(\D_{\mu})_{x+a_{\mu},x}
+(\D_{\mu})_{x+\tilde{a}+a_{\mu},x+\tilde{a}}(\tD)_{x+\tilde{a},x},
\label{link-ant-com}
\end{align}
which connect between  neighboring bosonic sites
via two different fermionic paths in a gauge covariant way.

\begin{figure}
\hfil
\begin{minipage}[b]{.4\linewidth}
\hfil \includegraphics[scale=.45]{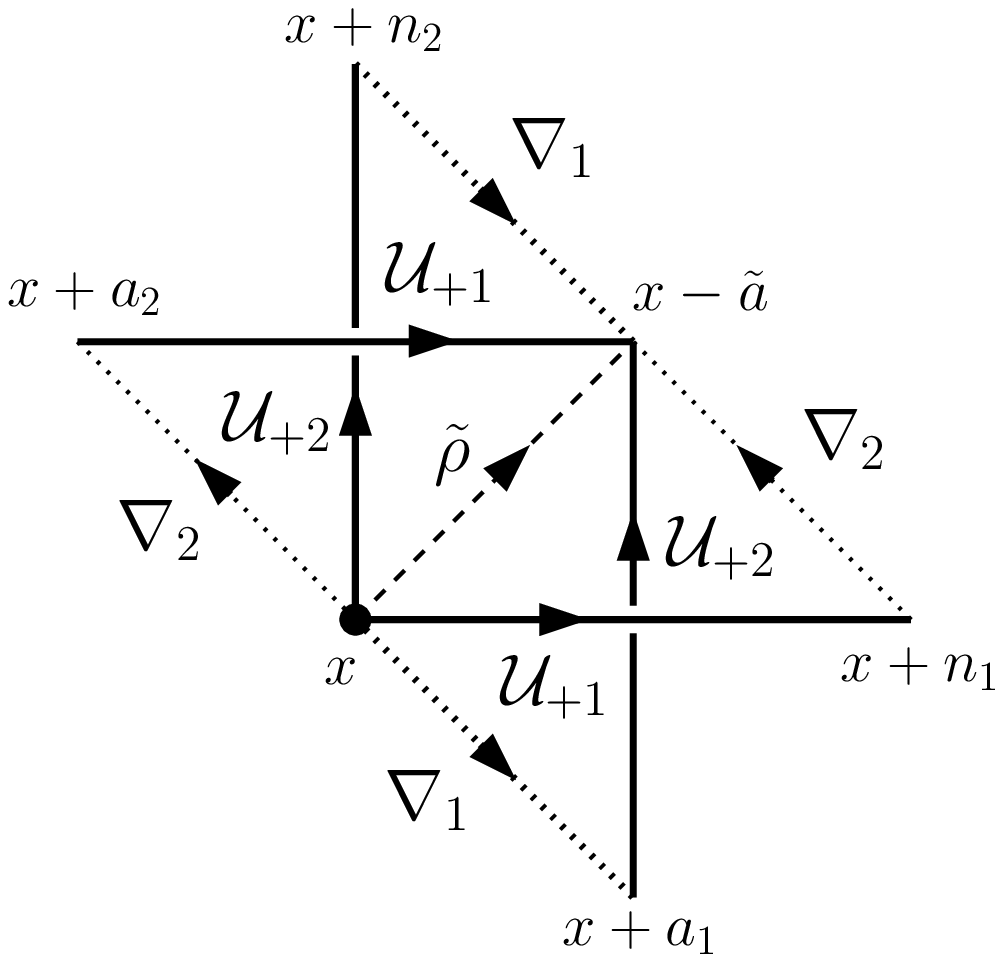}
\caption{Geometrical configuration around $\tilde{\rho}$}
\label{nabla_fermion}
\end{minipage}
\hfil
\begin{minipage}[b]{.55\linewidth}
\hfil \includegraphics[scale=.45]{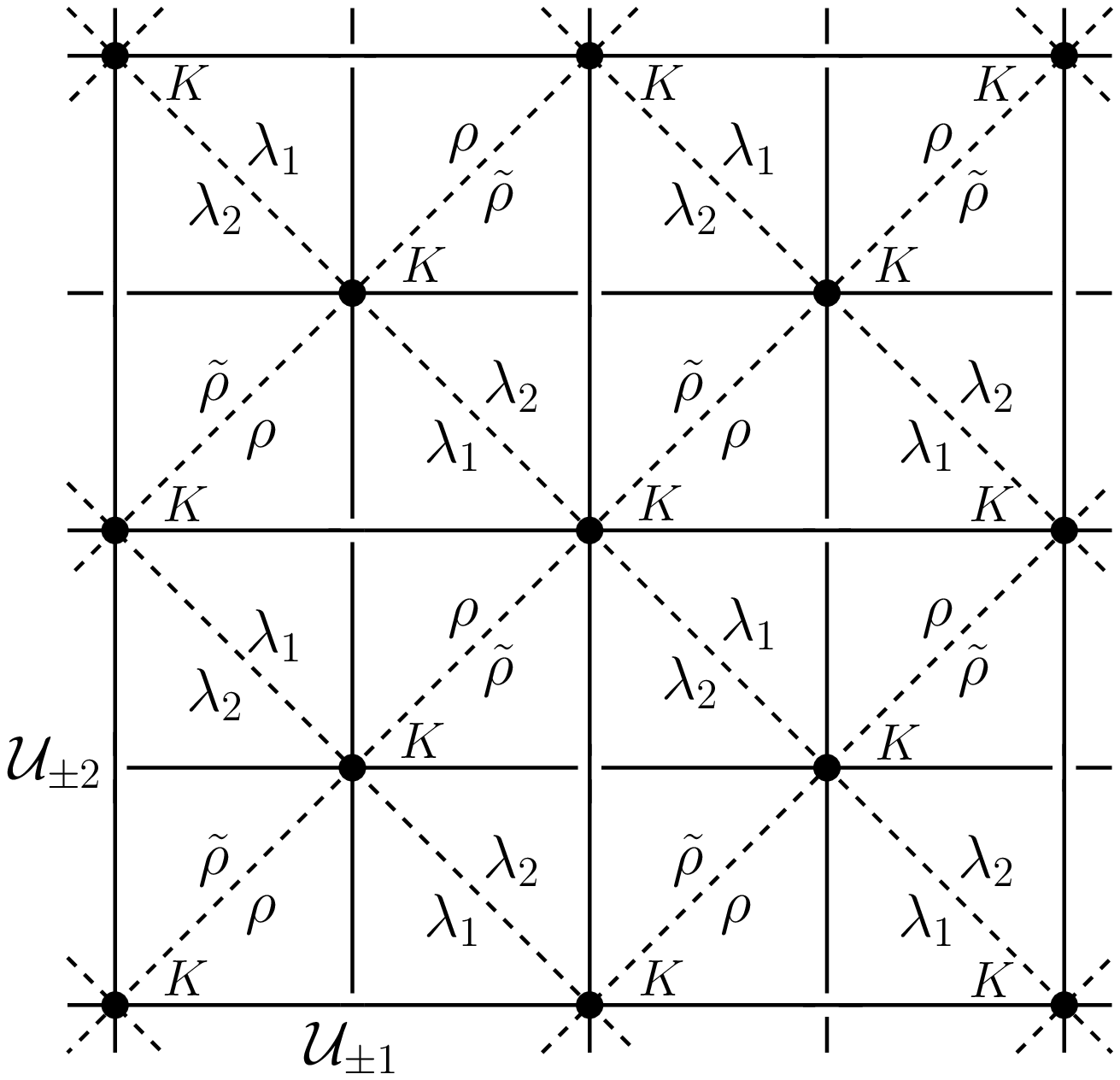}
\caption{Twisted $N=2$ SYM multiplet on a lattice}
\label{fig_fermion}
\end{minipage}
\hfil
\end{figure}

Jacobi identities of three fermionic link variables together with
the constrains (\ref{gauge1})--(\ref{gauge3}) give
\begin{eqnarray}
[\D,\U_{+\mu}]_{x+a+n_{\mu},x}\ 
=\ [\tD,\U_{-\mu}]_{x+\tilde{a}-n_{\mu},x} 
&=& 0,\label{lb1} \\[2pt]
[\D_{\mu},\U_{+\nu}]_{x+a_{\mu}+n_{\nu},x}
 + [\D_{\nu},\U_{+\mu}]_{x+a_{\nu}+n_{\mu},x} &=&0,\label{lb2} \\[2pt]
\epsilon_{\nu\lambda}[\D_{\mu},\U_{-\lambda}]_{x+a_{\mu}-n_{\lambda},x}
+\epsilon_{\mu\lambda}[\D_{\nu},\U_{-\lambda}]_{x+a_{\nu}-n_{\lambda},x}
 &=& 0, 
\label{lb3} \\[2pt]
\epsilon_{\mu\nu}[\D,\U_{-\nu}]_{x+a-n_{\nu},x}
+[\tD,\U_{+\mu}]_{x+\tilde{a}+n_{\mu},x} &=& 0,
\label{lb4} 
\end{eqnarray}
where again the l.h.s. should be understood as link commutators.
Note that within each equation (\ref{lb1})--(\ref{lb4}), 
the end points, which appear to be different,
are actually the same thanks to Eqs.(\ref{as}).
In agreement with the above relations, one may define the following 
non-vanishing fermionic link fields
\begin{eqnarray}
[\D_{\mu},\U_{+\nu}]_{x+a_{\mu}+n_{\nu},x} 
&\equiv& -\epsilon_{\mu\nu}(\tilde{\rho})_{x-\tilde{a},x}, \label{lf1}\\[2pt]
[\D_{\mu},\U_{-\nu}]_{x+a_{\mu}-n_{\nu},x} 
&\equiv& -\delta_{\mu\nu}(\rho)_{x-a,x}, \label{lf2} \\[2pt]
\epsilon_{\mu\nu}[\D,\U_{-\nu}]_{x+a-n_{\nu},x} 
= -[\tD,\U_{+\mu}]_{x+\tilde{a}+n_{\mu},x} &\equiv& 
-\epsilon_{\mu\nu}(\lambda_{\nu})_{x-a_{\nu},x},\label{lf3} 
\end{eqnarray}
which are $N=2$ twisted fermions on a lattice. 
The geometrical configuration underlying equation (\ref{lf1}) that  defines 
the fermionic link field $\tilde{\rho}_{x-\tilde{a},x}$ is shown in 
Fig.~\ref{nabla_fermion}. 
Due to the Jacobi identities  $\tilde{\rho}_{x-\tilde{a},x}$
is given by either $[\D_2,\U_{+1}]_{x+a_2+n_1,x}$  or $[\D_1,\U_{+2}]_{x+a_1+n_2,x}$.
With the first of the two choices, the fermionic link field 
$\tilde{\rho}_{x-\tilde{a},x}$
can be thought of as the difference of the two link paths
 $(\D_2)_{x-\tilde{a},x+n_1}(\U_{+1})_{x+n_1,x}$ and 
$(\U_{+1})_{x-\tilde{a},x+a_2}(\D_2)_{x+a_2,x}$. Same for the other choice.
Eqs.(\ref{as}) are obviously crucial for consistency.
Similar configurations could be easily found for the defining equations of
$\rho$  and $\lambda _\nu$.
These geometrical relations uniquely determine 
the relative locations of the fermionic link fields 
$\tilde{\rho}, \rho, \lambda_\mu$ and of the bosonic link fields $\U_{\pm\mu}$ 
on the lattice. This is shown, in the case of the symmetric choice of $a_A$,
in Fig.~\ref{fig_fermion}. 
It is important to realize that the bosonic links $\U_{\pm\mu}$ are 
composed of fermionic links as we can see in Eqs.(\ref{gauge1}) and 
(\ref{gauge2}) while a fermionic field such as 
$\tilde{\rho}, \rho, \lambda_\mu$ defined on the corresponding link are 
composed of the fermionic and bosonic link variables as we have just 
seen, and as a result, the lattice gauge covariance of the system is, 
by construction, manifest.    

\begin{table}
\renewcommand{\arraystretch}{1.3}
\renewcommand{\tabcolsep}{5pt}
\hfil
\begin{tabular}{c|c|c|c|c}
& $\D$ & $\tilde{\D}$ & $\D_{\mu}$ & $\U_{\pm\mu}$  \\ \hline
shift & $a$ &$\tilde{a}$ & $a_{\mu}$ & $\pm n_{\mu}$ 
\end{tabular}
\hfil
\begin{tabular}{c|c|c|c|c}
& $\rho$ & $\tilde{\rho}$ & $\lambda_{\mu}$ & $K$\\ \hline
shift & $-a$ & $-\tilde{a}$ & $-a_{\mu}$ & $0$
\end{tabular}
\caption{Shifts carried by link variables and fields}
\label{shifting_nature}
\end{table}

Jacobi identities of four fermionic link variables together with the
relations in (\ref{lf1})--(\ref{lf3}) lead to the following  relations:
{\allowdisplaybreaks
\begin{align}
\{\D,\tilde{\rho}\}_{x+a-\tilde{a},x} 
&\ =\  -\frac{i}{2}\epsilon_{\mu\nu}
[\U_{+\mu},\U_{+\nu}]_{x+n_{\mu}+n_{\nu},x},
\label{2lb5}\\
\{\tD,\rho\}_{x+\tilde{a}-a,x} &\ =\ 
+\frac{i}{2}\epsilon_{\mu\nu}
[\U_{-\mu},\U_{-\nu}]_{x-n_{\mu}-n_{\nu},x},
\label{2lb6}\\
\{\D,\rho\}_{x,x} &\ =\ -\frac{i}{2}[\U_{+\mu},\U_{-\mu}]_{x,x}-K_{x,x},
\label{2lb7}\\
\{\tD,\tilde{\rho}\}_{x,x} &\ =\
+\frac{i}{2}[\U_{+\mu},\U_{-\mu}]_{x,x}-K_{x,x},
\label{2lb8}\\
\{\D_{\mu},\lambda_{\nu}\}_{x+a_{\mu}-a_{\nu},x} &\ =\ 
-i[\U_{+\mu},\U_{-\nu}]_{x+n_{\mu}-n_{\nu},x}
+\delta_{\mu\nu}(K_{x,x}+\frac{i}{2}[\U_{+\rho},\U_{-\rho}]_{x,x}),
\label{21b9}\\[2pt]
\{\D_{\mu},\rho\}_{x+a_{\mu}-a,x} &\ =\ 
\{\D_{\mu},\tilde{\rho}\}_{x+a_{\mu}-\tilde{a},x}\ = \ 
\{\D,\lambda_{\mu}\}_{x+a-a_{\mu},x}\ = \ 
\{\tD,\lambda_{\mu}\}_{x+\tilde{a}-a_{\mu},x}\ = 0, 
\label{21b10} 
\end{align}
}
where $K_{x,x}=\frac{1}{2}\{\D_\mu,\lambda_\mu \}_{x,x}$ 
is an auxiliary field, being defined on a site, and ensures the off-shell 
closure of the twisted SUSY algebra on the lattice. 
The shift parameters defining the fermionic link fields and the 
auxiliary field are summarized in Table~\ref{shifting_nature}. 
It is interesting to note that all the relations derived from Jacobi 
identities in (\ref{2lb5})--(\ref{21b10}) have similar geometrical
interpretation
as the defining equation (\ref{lf1}) for $\tilde{\rho}$. 
For example the relation in (\ref{2lb5}) shows that 
$\{\D,\tilde{\rho}\}_{x+a-\tilde{a},x}$ denotes a doubly extended 
fermionic link pointing into the $a$ direction in the symmetric choice of the 
shift parameters ($a=-\tilde{a}$). As we can see from Fig.~\ref{fig_fermion}, 
this doubly 
extended fermionic link can be identified as two link paths in the bosonic link
commutator $[\U_{+1},\U_{+2}]_{x+n_1+n_2,x}$. The other relations have 
similar geometrical interpretations.   

SUSY transformation of twisted $N=2$ lattice gauge multiplets 
can be determined from the above relations resulting from the Jacobi identities via 
\begin{eqnarray}
(s_{A}\varphi)_{x+a_{\varphi}+a_A,x} = 
s_{A}(\varphi)_{x+a_{\varphi},x}\equiv
[\D_{A},\varphi\}_{x+a_{\varphi}+a_A,x},
\end{eqnarray}
where $(\varphi)_{x+a_{\varphi},x}$ denotes one of the component fields in 
$(\U_{\pm\mu},\rho,\tilde{\rho},\lambda_{\mu}.K)$. 
The results are summarized in Table~\ref{trans_lat}.
As a natural consequence of the formulation,
one can see that the $N=2$ SUSY algebra closes at off-shell 
(modulo gauge transformations) on a lattice,
\begin{eqnarray}
\{s,s_{\mu}\}(\varphi)_{x+a_{\varphi},x} 
&=& +i[\U_{+\mu},\varphi]_{x+a_{\varphi}+n_{\mu},x},
\label{uptogage1} \\[2pt]
\{\tilde{s},s_{\mu}\}(\varphi)_{x+a_{\varphi},x} &=& +i\epsilon_{\mu\nu}
[\U_{-\nu},\varphi]_{x+a_{\varphi}-n_{\nu},x},
\label{uptogage2} \\[2pt]
s^{2}(\varphi)_{x+a_{\varphi},x}&=&
\tilde{s}^{2}(\varphi)_{x+a_{\varphi},x} = 0,\\
\{s,\tilde{s}\}(\varphi)_{x+a_{\varphi},x}\ 
&=& \{s_{\mu},s_{\nu}\}(\varphi)_{x+a_{\varphi},x}\ =\ 0,
\end{eqnarray}
where again $\varphi$ denotes any component of the multiplet
$(\U_{\pm\mu},\rho,\tilde{\rho},\lambda_{\mu},K)$.

\begin{table}[t]
\renewcommand{\arraystretch}{1.4}
\renewcommand{\tabcolsep}{5pt}
\begin{tabular}{|c||c|c|c|}
\hline
& $s$ & $\tilde{s}$ & $s_{\mu}$ \\ \hline
$\U_{+\nu}$ & $0$ & $+\epsilon_{\nu\rho}\lambda_{\rho}$ &
 $-\epsilon_{\mu\nu}
\tilde{\rho}$\\
$\U_{-\nu}$ & $-\lambda_{\nu}$ & $0$ & $-\delta_{\mu\nu}\rho$ \\ 
$\lambda_{\nu}$ & $0$ & $0$ 
& $-i[\U_{+\mu},\U_{-\nu}]+\delta_{\mu\nu}(K+\frac{i}{2}[\U_{+\rho},
\U_{-\rho}])$\\
$\rho$ & $-\frac{i}{2}[\U_{+\rho},\U_{-\rho}]-K$ 
& $+\frac{i}{2}\epsilon_{\rho\sigma}[\U_{-\rho},\U_{-\sigma}]$
 & $0$ \\
$\tilde{\rho}$ & $-\frac{i}{2}\epsilon_{\rho\sigma}[\U_{+\rho},\U_{+\sigma}]$ 
& $+\frac{i}{2}[\U_{+\rho},\U_{-\rho}]-K$   & $0$ \\
$K$ & $+\frac{i}{2}[\U_{+\rho},\lambda_{\rho}]$ 
& $-\frac{i}{2}\epsilon_{\rho\sigma}[\U_{-\rho},\lambda_{\sigma}]$ &
$-\frac{i}{2}[\U_{+\mu},\rho]-\frac{i}{2}\epsilon_{\mu\nu}
[\U_{-\nu},\tilde{\rho}]$\\ \hline
\end{tabular}
\caption{SUSY transformation of $N=2$ lattice Super Yang-Mills multiplet}
\label{trans_lat}
\end{table}

\begin{figure}[t]
\hfil
\begin{minipage}[c]{35mm}
\hfil
\includegraphics[scale=.42]{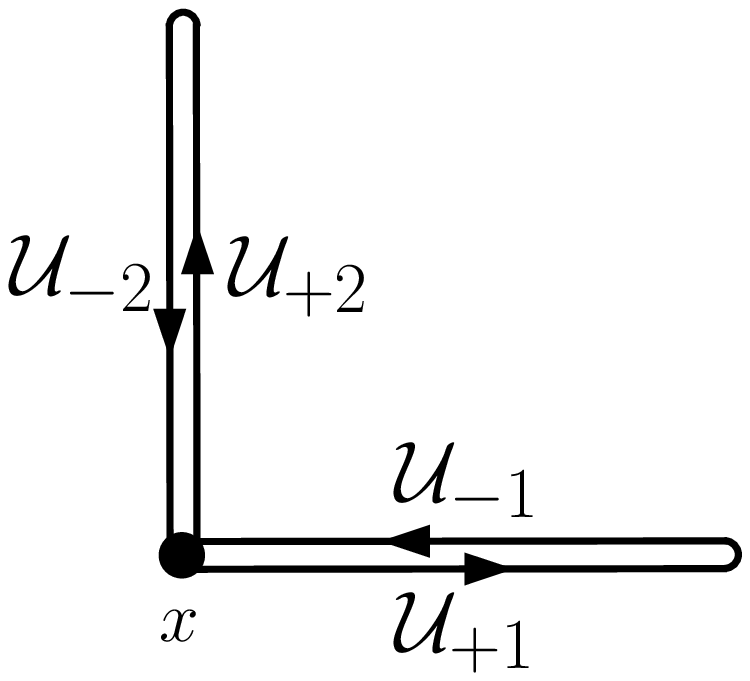}
\caption{Zero area loop on a link}
\label{zero-area-link}
\end{minipage}
\hfil
\begin{minipage}[c]{42mm}
\begin{center}
\includegraphics[scale=.42]{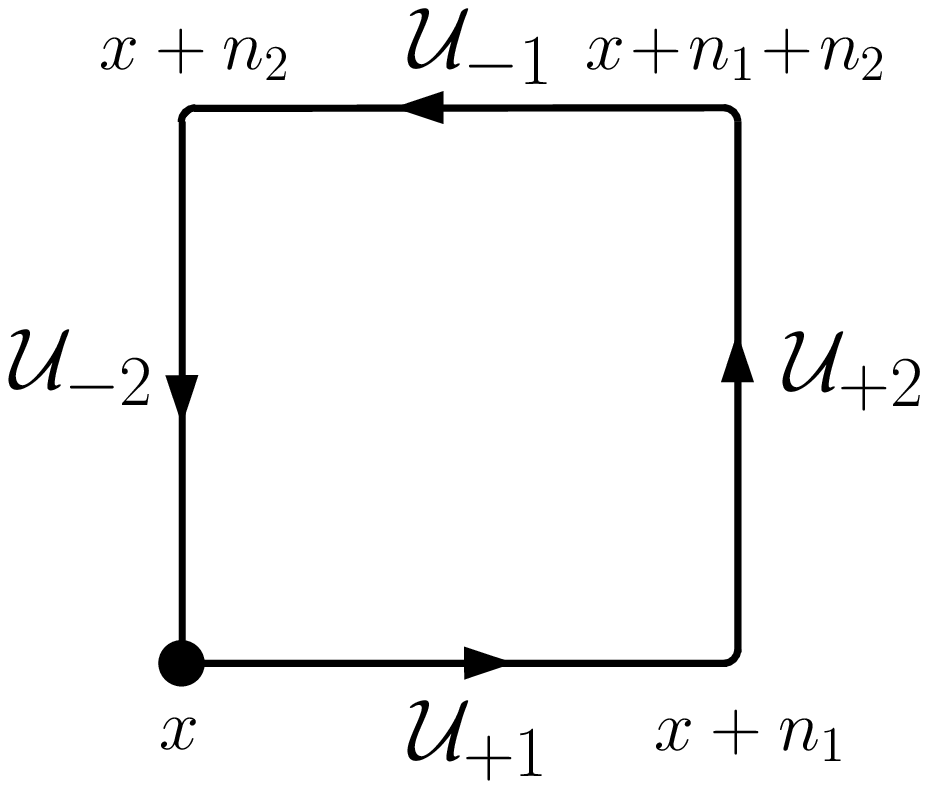}
\caption{Plaquette}
\label{plaquette}
\end{center}
\end{minipage}
\hfil
\begin{minipage}[c]{60mm}
\hfil
\includegraphics[scale=.42]{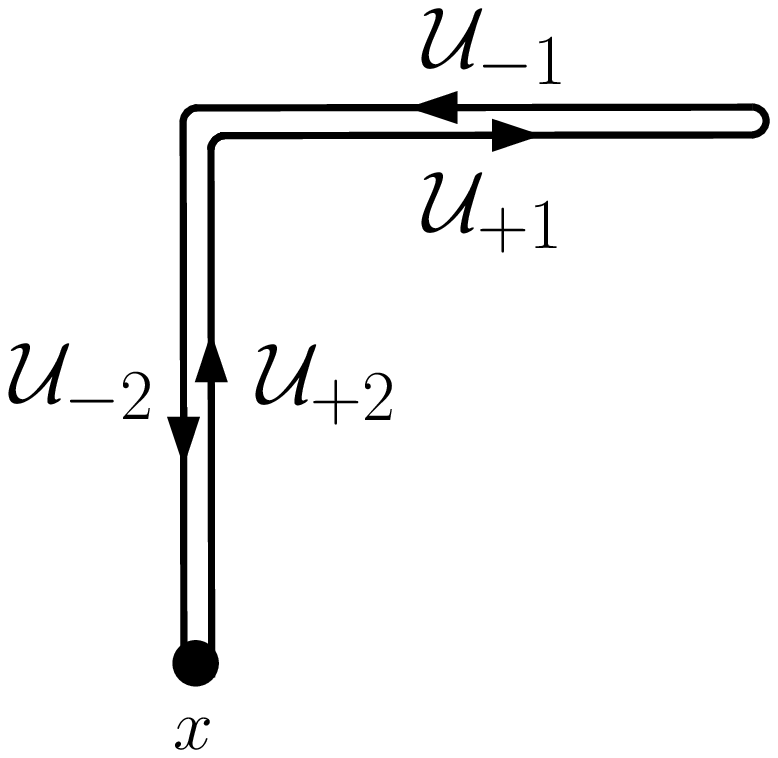}
\hspace{-1.5cm}
\includegraphics[scale=.42]{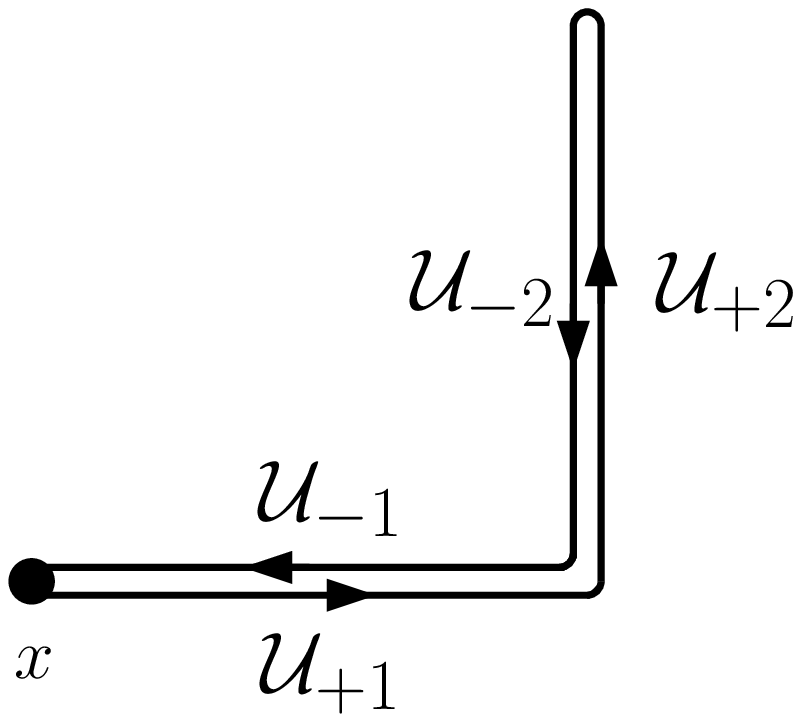}
\caption{Zero area loop extending over two links}
\label{non-unitary}
\end{minipage}
\vspace{5mm}

\hfil \includegraphics[scale=.45]{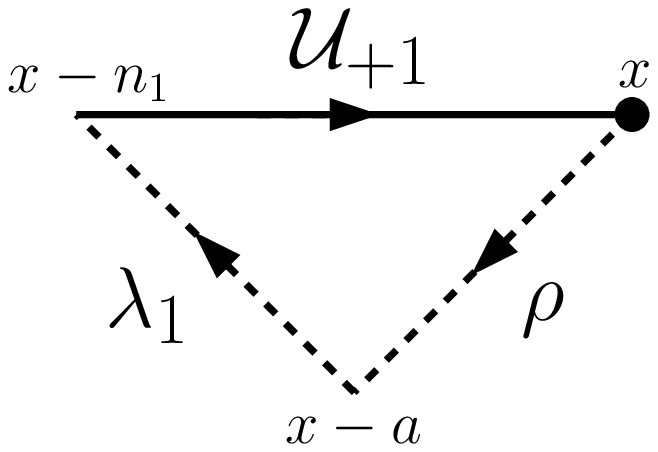}
\hfil \includegraphics[scale=.45]{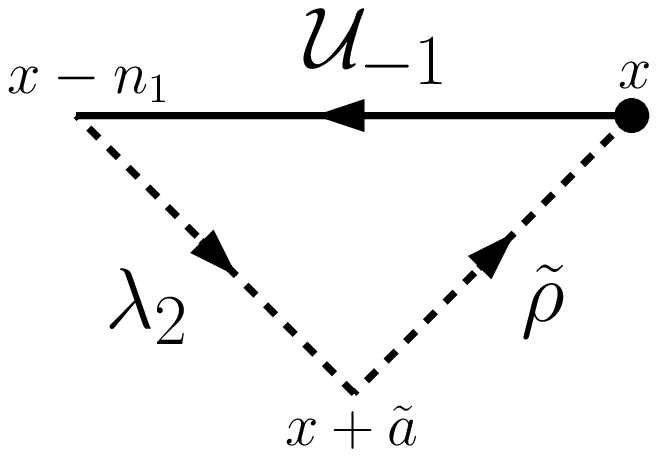}
\hfil \includegraphics[scale=.45]{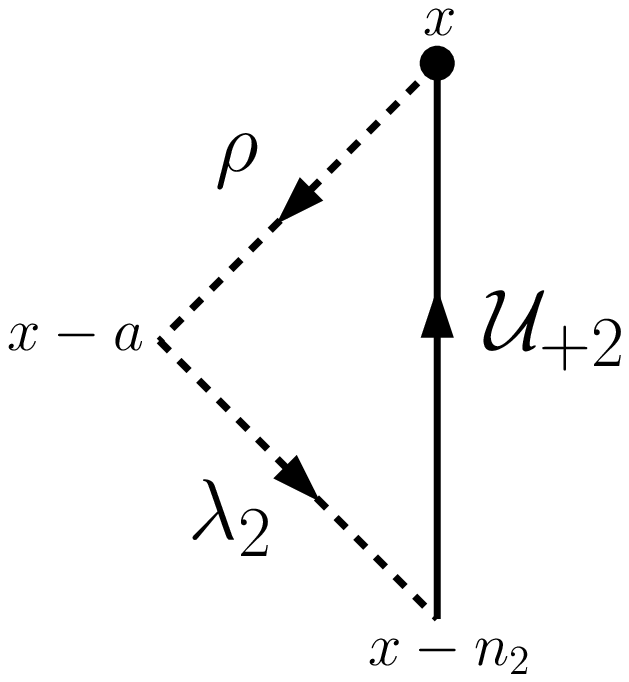}
\hfil \includegraphics[scale=.45]{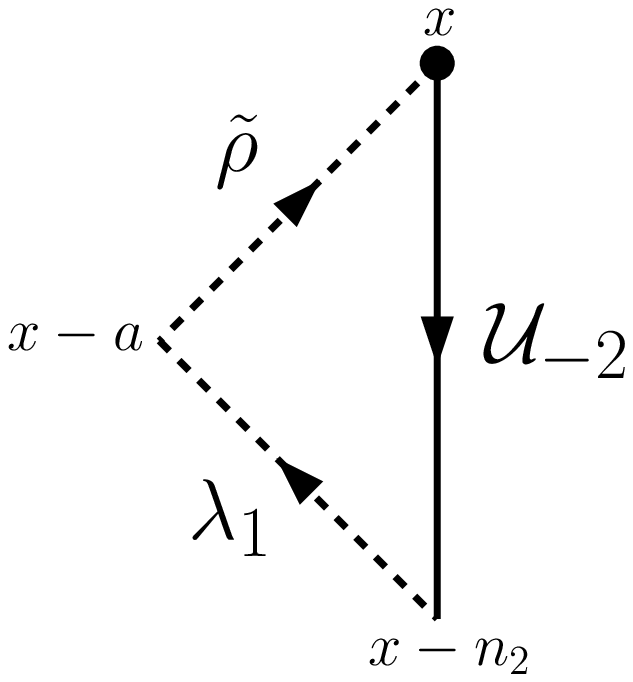}
\caption{Fermionic oriented loops} 
\label{fermionic-loop}
\end{figure}

The exact SUSY invariant action with respect to all the twisted 
supercharges on the lattice can be constructed by the 
successive operation of the twisted supercharge $s_{A}$ on
$\frac{1}{2}\,\U_{+\mu}\U_{-\mu}$,
\begin{eqnarray}
S &\equiv& \frac{1}{4}\sum_{x}\mathrm{Tr}\ 
s\tilde{s}\epsilon_{\mu\nu}s_{\mu}s_{\nu}\ \U_{+\mu}\U_{-\mu} 
\label{a1}\\
&=& \sum_{x}\mathrm{Tr}
\biggl[
\frac{1}{4}[\U_{+\mu},\U_{-\mu}]_{x,x}
[\U_{+\nu},\U_{-\nu}]_{x,x} +K_{x,x}^{2}\label{a2}\nonumber\\
&& -\frac{1}{4}\epsilon_{\mu\nu}\epsilon_{\rho\sigma}
[\U_{+\mu},\U_{+\nu}]_{x,x-n_{\mu}-n_{\nu}}
[\U_{-\rho},\U_{-\sigma}]_{x-n_{\rho}-n_{\sigma},x}
\label{a3}\nonumber\\
&& -i[\U_{+\mu},\lambda_{\mu}]_{x,x-a}(\rho)_{x-a,x}\ 
-\ i(\tilde{\rho})_{x,x+\tilde{a}}\epsilon_{\mu\nu}
[\U_{-\mu},\lambda_{\nu}]_{x+\tilde{a},x}
\label{a4}
\biggr],
\end{eqnarray}
where the summation over $x$ should cover the bosonic sites $(m_1,m_2)$
as well as fermionic sites $(m_1+\frac{1}{2},m_2+\frac{1}{2})$
($m_1,m_2$ : integers)
for the symmetric choice of the shift parameter 
$a_{A}$ (Fig.~\ref{symmetric_a}).
Due to this summation property, the order in the product 
of the supercharges
is irrelevant up to the gauge transformation of the form in (\ref{uptogage1}) and (\ref{uptogage2}).
Note that because of the exact form with respect to all the 
nilpotent supercharges, the action has, by construction, off-shell 
twisted SUSY invariance for all the supercharges.

This action has several interesting characteristics. 
Each term in the action consists of ``closed loop'' in a general 
sense. 
The first term in the action consists of zero area loops on a 
link as shown in Fig.~\ref{zero-area-link}. The second term is auxiliary 
field square defined on a site. The third term consists of a plaquette 
term which generates 
standard Yang-Mills action and zero area loops extending over two links 
shown in Fig.~\ref{plaquette} and Fig.~\ref{non-unitary}.
The fourth and fifth terms in the action consist of fermionic 
loops depicted in 
Fig.~\ref{fermionic-loop}
which have an opposite orientation.
It is important to recognize that the zero area loops in the action 
generate scalar terms since the non-unitary component in the link 
exponent of $\U_{\pm \mu}$ of (\ref{def-of-glink}) remains in the term like 
$\U_{+\mu}\U_{-\mu}$. 

The closed loop nature of all terms in the action 
ensures the manifest gauge invariance. Notice that 
this invariance has a deep relation to the arguments 
of the Leibniz rule on the lattice since 
the closure nature of each loop can be traced back to
the shift-less property of $s\tilde{s}s_{1}s_{2}$
which is the direct consequence of the consistency condition
for the Leibniz rule on a lattice, namely,
the vanishing sum of $a_{A}$ in (\ref{as}).

The naive continuum limit of the action (\ref{a4}) can be taken 
through the expansion of gauge link variables (\ref{def-of-glink}),
\begin{eqnarray}
(\U_{\pm\mu})_{x\pm n_{\mu},x} 
= (e^{\pm i(A_{\mu}\pm i\phi^{(\mu)})})_{x\pm n_{\mu},x}
= (1 \pm i(A_{\mu}\pm i\phi^{(\mu)})+\cdots)_{x\pm n_{\mu},x},
\end{eqnarray}
where we identify that $A_\mu(x+\frac{n_\mu}{2})$ and 
$\phi^{(\mu)}(x+\frac{n_\mu}{2})$ are located on a middle of link. 
Inserting the expansion and using some trace properties, we obtain 
a continuum action 
\begin{eqnarray}
S\rightarrow S_{cont} 
&=& \int d^{2}x\mathrm{Tr}\biggl[
\frac{1}{2}
F_{\mu\nu}F_{\mu\nu}
+[\cD_{{\nu}},\phi^{(\mu)}][\cD_{{\nu}},\phi^{(\mu)}]+K^{2}
\nonumber\\
&&-\frac{1}{2}[\phi^{(\mu)},\phi^{(\nu)}][\phi^{(\mu)},\phi^{(\nu)}]
 +i[\cD_{{\mu}},\lambda_{\mu}]\rho\ 
-\ i\tilde{\rho}\epsilon_{\mu\nu}[\cD_{{\mu}},\lambda_{\nu}]
\nonumber \\
&&
 +i[\phi^{(\mu)},\lambda_{\mu}]\rho\ +\ i\tilde{\rho}\epsilon_{\mu\nu}
[\phi^{(\mu)},\lambda_{\nu}]
\biggr],
\label{action_cont}
\end{eqnarray}
where $F_{\mu\nu}\equiv i[\cD_{\mu},\cD_{\nu}]$ denotes
ordinary gauge field strength with
$\cD_{\mu}\equiv \partial_{\mu}-iA_{\mu}$ 
while $\phi^{(\mu)}(\mu=1,2)$,
as mentioned above,
represent the two independent scalar fields in the $N=D=2$ twisted SYM
multiplet.
In fact, (\ref{action_cont}) can be shown to have complete 
agreement with the continuum construction of $N=D=2$ twisted SYM.

One can as well take the continuum limit
of the lattice SUSY transformation of the component fields. For example,
the lattice SUSY transformations: 
$ \tilde{s}\ \U_{+\nu}=+\epsilon_{\nu\rho}\lambda_{\rho} $
and $\tilde{s}\ \U_{-\nu}=0 $, respectively, lead 
\begin{eqnarray}
\tilde{s}\ (1+i(A_{\nu}+i\phi^{(\nu)})+\cdots)
= \epsilon_{\nu\rho}\lambda_{\rho},
\hspace{30pt}
\tilde{s}\ (1-i(A_{\nu}-i\phi^{(\nu)})+\cdots)=0,
\end{eqnarray}
from which we find their continuum counterparts as
\begin{eqnarray}
\tilde{s}A_{\nu}=-\frac{i}{2}\epsilon_{\nu\rho}\lambda_{\rho},\hspace{30pt}
\tilde{s}\phi^{(\nu)}=-\frac{1}{2}\epsilon_{\nu\rho}\lambda_{\rho}.
\end{eqnarray}
Other SUSY transformation of the component fields in the continuum limit 
can also be uniquely determined in the same manner. 
Table~\ref{trans_cont} summarizes the twisted SUSY transformation 
in the continuum limit for $A_{\nu}$ 
and $\phi^{(\nu)}$, which completely agrees with those of the continuum 
construction. 
The SUSY transformation of the fermionic fields and auxiliary field can 
be just read off from the lattice counterpart of Table~\ref{trans_lat} 
by simply replacing 
$\U_{\pm\mu} \rightarrow \mp(\partial_{\mu} -iA_\mu \pm\phi^{(\mu)})$.

\begin{table}
\begin{center}
\renewcommand{\arraystretch}{1.5}
\renewcommand{\tabcolsep}{10pt}
\begin{tabular}{|c||c|c|c|}
\hline
& $s$ & $\tilde{s}$ & $s_{\mu}$ \\ \hline
$A_{\nu}$ & $-\frac{i}{2}\lambda_{\nu}$ 
& $-\frac{i}{2}\epsilon_{\nu\rho}\lambda_{\rho}$ &
$+\frac{i}{2}\epsilon_{\mu\nu}\tilde{\rho}-\frac{i}{2}\delta_{\mu\nu}\rho$ \\ 
$\phi^{(\nu)}$ & $+\frac{1}{2}\lambda_{\nu}$ 
& $-\frac{1}{2}\epsilon_{\nu\rho}\lambda_{\rho}$ 
& $+\frac{1}{2}\epsilon_{\mu\nu}\tilde{\rho}
+\frac{1}{2}\delta_{\mu\nu}\rho$ \\ \hline
\end{tabular}
\caption{SUSY transformation laws for $A_{\nu}$ and $\phi^{(\nu)}$}
\label{trans_cont}
\end{center}
\end{table}

\section{Discussions}

Lattice actions for SYM theories in which the scalar part 
of twisted supersymmetry is exactly preserved on the lattice were 
proposed by several authors\cite{twist-lattsusy}. 
In particular the connection between the Dirac-K\"ahler mechanism and 
twisted SUSY for SYM was also stressed in \cite{twist-lattsusy2}. 
It was argued that scalar part of the twisted SUSY invariance 
assures the recovery of the full SUSY invariance in the continuum limit. 
In our approach all supersymmetries of the twisted $N=2$ SUSY are exactly
realized in the lattice action for two dimensional $N=2$ SYM, as a natural 
consequence of the Dirac-K\"ahler twisting mechanism for the supercharges.
We have introduced a lattice where the anti-commutators of the SUSY algebra
and the Jacobi identities can be read off as geometric relations amongst 
the elements of the lattice and we have shown that the link nature of the 
difference operator and of the super covariant derivatives is in fact equivalent
(but conceptually much more powerful) to the ``mild non-commutativity''
introduced in our previous paper.
  
It is also worth mentioning that $N=D=2$ SYM lattice action, with just one
exact supersymmetry, was also obtained by the orbifold-deconstruction 
method in \cite{deconstruction}. Although the connection between this method
and our approach is unclear, it is interesting that the lattice in 
\cite{deconstruction} appears to be a degenerate case of our lattice,
namely the one corresponding to the choice $a=(0,0)$ of the 
shift parameter. 
 
The details of the present formulation together with the connection with 
other approaches will be given in a separate publication\cite{DKKN2}. 

\subsection*{Acknowledgments}
We would like to thank to J.~Kato, A.~Miyake and J.~Saito for useful discussions.
This work is supported in part by Japanese Ministry of Education,
Science, Sports and Culture under the grant number 13135201 and also by
INFN research funds.


\end{document}